\documentclass[aps,prl,twocolumn,superscriptaddress]{revtex4-2}
\usepackage{amssymb,color}
\usepackage{graphicx}
\usepackage{slashed}
\usepackage{amsmath}
\usepackage{eso-pic}
\def\ba{\begin{eqnarray}}
\def\ea{\end{eqnarray}}
\newcommand{\nl}{\nonumber \\ }
\newcommand{\nn}{\nonumber }

\allowdisplaybreaks[3]

\def \bal#1\eal  {\begin{align} #1 \end{align}}
\def\({\left(}
\def\){\right)}
\def\[{\left[}
\def\]{\right]}
\def\<{\langle}
\def\>{\rangle}
\def\d{\mathrm{d}}

\newcommand{\f}[2]{\frac{#1}{#2}}
\newcommand{\bim} {\begin{itemize}[noitemsep]}

\newcommand{\eim}{\end{itemize}}
\newcommand{\be} {\begin{equation}}
\newcommand{\ee} {\end{equation}}
\newcommand{\bc}{\begin{center}}
\newcommand{\ec}{\end{center}}

\begin{document}
\AddToShipoutPictureFG*{ 
     \AtPageUpperLeft{\put(-78,-40){\makebox[\paperwidth][r]{USTC-ICTS/PCFT-23-30}}}  
    }

\title{Bi-gravity Portal Dark Matter}

\author{Qing Chen}
\email[]{cqpb@ustc.edu.cn}
\affiliation{Interdisciplinary Center for Theoretical Study, University of Science and Technology of China, Hefei, Anhui 230026, China}
\affiliation{Peng Huanwu Center for Fundamental Theory, Hefei, Anhui 230026, China}
\author{Shuang-Yong Zhou}
\email[]{zhoushy@ustc.edu.cn}
\affiliation{Interdisciplinary Center for Theoretical Study, University of Science and Technology of China, Hefei, Anhui 230026, China}
\affiliation{Peng Huanwu Center for Fundamental Theory, Hefei, Anhui 230026, China}
\affiliation{Theoretical Physics, Blackett Laboratory, Imperial College, London, SW7 2AZ, UK}

\date{\today}

\begin{abstract}
We consider a model where the interaction between dark matter and the Standard Model particles are mediated by a ghost-free bi-gravity portal. The bi-gravity model invokes a massive spin-2 particle coupled to the usual massless graviton as well as generic bi-metric matter couplings. The cross-sections for dark matter direct detection are computed and confronted with the experimental bounds. The presence of the massive spin-2 mediator resolves the core-cusp problem, which in turn significantly constrains the dark matter coupling in such a bi-gravity theory. Yet, there remains a window of the parameter space where the model can be tested in the upcoming direct detection experiments such as XENONnT and PandaX-30T. The model also predicts a reheating temperature of the order of $10^6$\,GeV.
\end{abstract}

\maketitle

\section{Introduction}
Null results in the non-gravitational experiments, which have achieved an extraordinary level of precision\,\cite{LZ:2022lsv, PandaX-4T:2021bab, XENON:2018voc, Fermi-LAT:2019lyf, ANTARES:2013vvr, PerezAdan:2023rsl}, leave dark matter (DM)\,\cite{Bertone:2004pz, Feng:2010gw, ParticleDataGroup:2022pth} one of the most outstanding problems in particle physics and cosmology. If particle-like dark matter exists, its interaction with the Standard Model (SM) sector must be sufficiently weak so that we have not observed it directly.
In this letter, we investigate a scenario where such a feeble interaction portal is provided by spin-2 mediators.  Indeed, in Einstein's gravity, gravitational force mediated by the massless spin-2 graviton is much weaker than the other 3 known fundamental interactions. Thanks to its universal coupling, this graviton portal allows us to identify the existence of dark matter\,\cite{1933AcHPh...6..110Z, Rubin:1985ze,Tyson:1998vp,Inada:2003vc,Jee:2007nx,Clowe:2006eq}. If it is the only SM-DM portal, dark matter can not be observed in direct detection experiments. We consider a model where there is an additional massive spin-2 particle which acts as the main SM-DM portal at short range, that is, a bi-gravity model with composite matter couplings. In the bi-gravity construction, the spin-2 gauge group (two copies of diffeomorphism invariance) necessarily breaks down to a subgroup $\mathrm{Diff}_L\times \mathrm{Diff}_R\to \mathrm{Diff}_V$\,\cite{Arkani-Hamed:2002bjr}, which in some sense is analogous to the spontaneous symmetry breaking in the spin-1 gauge group of the SM electroweak sector. 
Consequently, after symmetry breaking, the massless spin-2 particle is the familiar Einstein graviton.
The massive spin-2 particle mediates a short-range fifth force, which is well within the bounds set by the current gravity tests.
 
Our model is different from the spin-2 dark matter models where the dark matter particles themselves are of a spin-2 nature\,\cite{Aoki:2016zgp,Babichev:2016hir,Babichev:2016bxi}, where the interactions with the SM sector are too weak to produce any direct detection signals. Interactions between dark matter and the SM particles via channels of a non-trivial gravitational nature have been previously explored.
Specifically, it has been considered in the context of the Randall-Sundrum extra-dimension models, 
where dark matter interacts with the SM matter via Kaluza-Klein resonances\,\cite{Lee:2013bua, Rueter:2017nbk, Goyal:2019vsw, Folgado_2020}. This is different from the sharp-mass bi-gravity portal we are considering. Indeed, our model builds up on the recent developments in constructing consistent massive gravity/bi-gravity models \cite{deRham:2010kj, Hassan:2011zd}(see \cite{deRham:2014zqa, Schmidt-May:2015vnx} for a review), especially the consistent forms of bi-metric couplings to matter \cite{deRham:2014zqa, Huang:2015yga}. On the other hand, the DM relic abundance via $s$-channel spin-2 portal effective interactions in the early universe were computed in\,\cite{Bernal:2018qlk}, and in Ref\,\cite{Kraml:2017atm} the LHC signals involving $s$-channel spin-2 mediators were examined within a simplified phenomenological framework. In contrast, our investigation delves into $t$-channel spin-2 mediators in the context of dark matter direct detection, beginning with a comprehensive Lagrangian featuring predefined theory parameters. We will see that the parameter choices are more limited starting from the Lagrangian, and if such a massive spin-2 portal exists, a specific region of the bi-gravity theory space is preferred in order to resolve the small-scale anomaly in the DM astronomy, the core-cusp problem\,\cite{Moore:1994yx,Flores:1994gz,Navarro:1995iw,Navarro:1996gj}, which can be tested in the next generation direct detection experiments. Also, by estimating the correct dark matter relic abundance in this bi-gravity model within the viable parameter region, we can predict a reheating temperature in the early Universe.

\section{The model}

As mentioned, we consider a gravitational sector where the conventional massless graviton is accompanied by a massive counterpart, {\it i.e.,} a bi-gravity model. Gravity models with a massive graviton were traditionally known to be plagued by various theoretical problems such as the Boulware-Deser ghost\,\cite{Boulware:1972yco}. Recent years have seen significant advances in constructing and understanding ghost-free massive gravity models, especially after the discovery of the dRGT graviton potential\,\cite{deRham:2010kj}. Extensive literature has focused on the possibility of a Hubble-scale mass for the spin-2 particle, and in such a scenario the model could potentially explain the late time cosmic acceleration or the dark energy problem (see\,\cite{deRham:2014zqa, deRham:2016nuf} for a review). Nevertheless, when detached from the context of the dark energy problem, the mass of the spin-2 particle can be significantly larger. In this paper, we will explore whether such a scenario can accommodate an interaction portal to dark matter that can be observed in the upcoming direct detection experiments. 

Let us first specify the model. A dRGT-type bi-gravity Lagrangian\,\cite{Hassan:2011zd, Schmidt-May:2015vnx} is given by 
\bal
S_{\rm bg} &=\frac{M_{\mathrm{pl}}^2}{2}\int \d^4x\sqrt{|g|}\,R[g]+\frac{M_f^2}{2}\int \d^4x\sqrt{|f|}\,R[f]\nn
\\
&\quad ~ +\frac{M_{f}^2m^2}{4}\int \d^4x\sqrt{|f|}\,\sum_{n=0}^4\alpha_n U_n(K[g,f])\,,
\label{bi}
\eal
where $R[g]$ and $R[f]$ are the Ricci scalars built out of metric $g_{\mu\nu}$ and $f_{\mu\nu}$ respectively, $M_{\mathrm{pl}}$ is the reduced Planck mass and $M_f$ is another mass scale chosen to be $M_f\ll M_{\mathrm{pl}}$, which will allow strong matter couplings for the massive spin-2 mode. $\alpha_n$'s are real coefficients, and we choose $\alpha_1=0$ and $\alpha_2=2$ to eliminate the cosmological constant and the tadpole term. The unique dRGT potential terms, which completely remove the Boulware-Deser ghost\,\cite{deRham:2010kj, Hassan:2011hr}, take the form 
\be
U_n(K)=K^{\mu_1}_{[\mu_1}K^{\mu_2}_{\mu_2}\dots K^{\mu_n}_{\mu_n]},~~ K^{\mu}_{\nu}\equiv \delta^{\mu}_{\nu} - \sqrt{f^{-1}g}\big|^\mu_\nu ,
\ee
where ${}_{[~]}$ denotes anti-symmetrization of the indices and the square root of the matrix $f^{-1}g$ ($f^{-1}$ being the matrix of the inverse metric $f^{\mu\nu}$) is defined by taking its principal branch. From the perspective of symmetries, the two copies of diffeomorphism invariance in bi-gravity are broken down by the spin-2 potential terms to a single diagonal subgroup. We assume that the Standard Model and dark matter particles couple to both metrics but differently\,\footnote{Note that for the coupling to a fermionic field in curved space, the below metric formulation is schematic, and vierbeins and spin-connections should be used instead. For example, for a fermion $\psi$, the action goes like $\int d^4x \,\mathrm{det}\left(e^a_\mu\right)\left( \f12 i \bar{\psi} \gamma^a e^\mu_a {D}_{-\mu} \psi -m\bar{\psi}\psi\right)$, where $\gamma^a$ are the usual flat-space gamma matrices, $e^\mu_a$ is the vierbein and ${D}_{-\mu}\equiv D _\mu- \overleftarrow{D}_\mu$ with ${D}_\mu=\partial_\mu-ig_s A_\mu+\frac{1}{8}\omega_\mu^{ab} [\gamma_a, \,\gamma_b]$ is defined with spin connection $\omega_\mu^{ab}$, $a,b,...$ being the flat space indices on the local Lorentz frame.}, 
\be
S_{\rm M} =\int d^4x\sqrt{|g^{\mathrm{eff}}|}\, \mathcal{L}_{\mathrm{SM}}+\int d^4x\sqrt{| f^{\mathrm{eff}}|}\, \mathcal{L}_{\mathrm{DM}}\,,
\label{matterL}
\ee
with
\bal
g^{\mathrm{eff}}_{\mu\nu}&=\alpha^2 f_{\mu\nu}+2\alpha\beta f_{\mu\rho} \sqrt{f^{-1}g}\big|^\rho_\nu+\beta^2 g_{\mu\nu}\,,
\\
f^{\mathrm{eff}}_{\mu\nu}&=\alpha^{\prime2} f_{\mu\nu}+2\alpha^\prime\beta^\prime f_{\mu\rho} \sqrt{f^{-1}g}\big|^\rho_\nu+\beta^{\prime 2} g_{\mu\nu}\,,
\eal
The effective/composite metrics $g^{\rm eff}_{\mu\nu}$ and $f^{\rm eff}_{\mu\nu}$ are constructed in this way to avoid re-introducing the Boulware-Deser ghost below the EFT cutoff either classically or under loop corrections, thanks to utilizing the dRGT matrix $K^\mu_\nu$ or $\sqrt{f^{-1}g}$ in these effective metrics\,\cite{deRham:2014zqa}. In fact, it can be shown that this choice of the effective metric is the only one to achieve this avoidance\,\cite{Huang:2015yga}. 

After two copies of diffeomorphism in Eq.\,(\ref{matterL}) breaking into one, we obtain a massive mode $H_{\mu\nu}$ and a massless mode $h_{\mu\nu}$. At linear order, we have 
\begin{align}
&g^{\mathrm{eff}}_{\mu\nu}=\eta_{\mu\nu}+\kappa \frac{H_{\mu\nu}}{M_{\mathrm{pl}}}
+\xi_r \frac{h_{\mu\nu}}{M_{\mathrm{pl}}}\,,\label{g_eff}
\\
&f^{\mathrm{eff}}_{\mu\nu}=\eta_{\mu\nu}+\kappa^\prime \frac{H_{\mu\nu}}{M_{\mathrm{pl}}}
+\xi_r \frac{h_{\mu\nu}}{M_{\mathrm{pl}}}\,,\label{f_eff}
\end{align}
with $\xi_r\equiv\left(1+r^2\right)^{-1/2}$, $\kappa\equiv \xi_r\left[1-(\beta/\xi_r)^2\right]/r$ and $\kappa^\prime\equiv \xi_r\left[1-(\beta^{\prime}/\xi_r)^2\right]/r$, where $\eta_{\mu\nu}=\mathrm{diag}(1,\,-1,\,-1,\,-1)$, we have defined a ratio $r\equiv M_f/M_{\mathrm{pl}}$ and chosen the normalization $\left(\alpha+\beta\right)^2 = \left(\alpha^\prime+\beta^\prime\right)^2=1$. The mass of the massive spin-2 mode is given by
\be
m_{\mathrm{eff}}=m/\xi_r.
\ee

\section{Bi-gravity portal and direct detection cross section}

\begin{figure}[t]
\includegraphics[width=\linewidth]{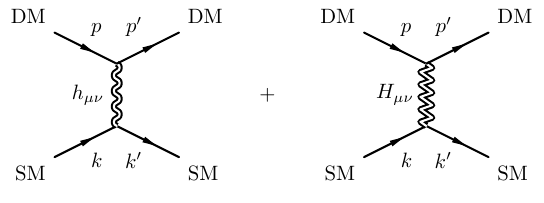}
\caption{SM (nucleon) and DM scattering via a massless portal ($h_{\mu\nu}$) and a massive portal ($H_{\mu\nu}$).} 
\label{sm_dm_diag}
\end{figure}

In this model, the Standard Model and dark matter particles can interact via spin-2 portals. That is, we have the process $\mathrm{SM}+\mathrm{DM}\to \mathrm{SM}+\mathrm{DM}$ as shown in Fig.~\ref{sm_dm_diag}, where the SM particles are taken to be nucleons. This is the primary signal channel in dark matter direct detection experiments. The nucleon's compositeness is negligible for the interaction we are considering here, and it will be treated as a spin-1/2 massive point-like particle. 
Let us compute the cross section of this process under the assumption $M_f\ll M_{\mathrm{pl}}$. In this scenario, the amplitude is proportional to $M_{\mathrm{pl}}^{-2}$ via the massless spin-2 portal, the same as that in general relativity, but is proportional to $(1-\beta^{\prime 2})(1-\beta^2)r^{-2}M_{\mathrm{pl}}^{-2}$ via the massive spin-2 portal, which can be significantly different.

To have observable signals in direct detection experiments, we consider an elastic process for $\mathrm{SM}+\mathrm{DM}\to \mathrm{SM}+\mathrm{DM}$ and assume DM particles are much heavier than the nucleon, for which we only need to include the $t$-channel contribution, as shown in Fig.~\ref{sm_dm_diag}. In this case, the amplitude for the scattering between a spin-$s$ DM particle and a nucleon is given by 
\bal
i\mathcal{M}_{(s)}&=\sum_{i=h,\,H}g_i T^{\mu\nu}_{(s)}  P^i_{\mu\nu\alpha\beta}
\bar{u}_N(k^\prime)\left(-\frac{i}{8}\right)\Big[(k^\alpha+k^{\prime\alpha})\gamma^\beta 
\nl
&\hspace{-25pt} + (k^\beta+k^{\prime\beta})\gamma^\alpha -2\eta^{\alpha\beta}(\slashed{k}+\slashed{k}^\prime)
+4m_N \eta^{\alpha\beta}\Big]u_N(k)\,,\label{eq:amp}
\eal
where the $T^{\mu\nu}_{(s)}$ tensor for a scalar ($s$=0) DM particle is
\ba
T^{\mu\nu}_{(0)}=-\frac{i}{2}\left[p^\mu p^{\prime\nu}+p^\nu p^{\prime\mu}+\eta^{\mu\nu}\left(M^2-p\cdot p^\prime\right)\right]\,,
\ea
and that of a fermionic ($s=1/2$) DM particle is
\ba
T^{\mu\nu}_{(\frac12)}&&=\bar{u}_{\chi}(p^\prime)\left(-\frac{i}{8}\right)\Big[(p^\mu+p^{\prime\mu})\gamma^\nu+(p^\nu+p^{\prime\nu})\gamma^\mu
\nl
&&-2\eta^{\mu\nu}(\slashed{p}+\slashed{p}^\prime)
+4M \eta^{\mu\nu}\Big]u_\chi(p)\,, 
\ea
$m_N$ and $M$ being the nucleon mass and the DM mass respectively.
For the massless portal, the effective coupling is $g_h = \xi_r^2M_{\mathrm{pl}}^{-2}$ and the massless spin-2 graviton propagator is
\be
P^h_{\mu\nu\alpha\beta}=\frac{1}{2}\frac{-i\left({\eta}_{\mu\alpha}{\eta}_{\nu\beta}+{\eta}_{\mu\beta}{\eta}_{\nu\alpha}
-{\eta}_{\mu\nu}{\eta}_{\alpha\beta}\right)}{q^2+i\epsilon}\,,
\ee
where $q^\mu=p^{\prime \mu}-p^\mu$, while, for the massive portal, the effective coupling is $g_H\equiv \kappa\kappa^\prime M_{\mathrm{pl}}^{-2}$ and the massive spin-2 particle's propagator is
\be
P^H_{\mu\nu\alpha\beta}=\frac{1}{2}\frac{-i\left(\tilde{\eta}_{\mu\alpha}\tilde{\eta}_{\nu\beta}+\tilde{\eta}_{\mu\beta}\tilde{\eta}_{\nu\alpha}
-\frac{2}{3}\tilde{\eta}_{\mu\nu}\tilde{\eta}_{\alpha\beta}\right)}{q^2-m_{\mathrm{eff}}^2+i\epsilon}\,,
\ee
where $\tilde{\eta}_{\mu\nu}=-{\eta}_{\mu\nu}+q_{\mu}q_\nu/m^2_{\mathrm{eff}}$. 

In the center-of-mass frame, the 4-momenta can be written as
$p^\mu=(E_\chi, \mathbf{p})$,
$k^\mu=(E_N, -\mathbf{p})$,
$p^\prime{}^\mu=(E_\chi, \mathbf{p}+\mathbf{q})$,
$k^\prime{}^\mu=(E_N, -\mathbf{p}-\mathbf{q})$. Due to the elasticity of the scattering, we have $|\mathbf{p}|=|\mathbf{p}+\mathbf{q}|$ and $q^2=-\mathbf{q}^2=2|\mathbf{p}|^2(\cos{\theta}-1)$, with $\theta$ being the scattering angle between $p^\mu$ and $p'{}^\mu$. For DM particles in the non-relativistic regime with velocity $v\sim 10^{-3}$, we can approximate $E_\chi$ and $E_N$ with $M$ and $m_N$ respectively. As mentioned, here we shall restrict ourselves to the scenario of relative heavier dark matter particles $M\gg m_N$, as it results in more sizable cross sections. With these approximations, the DM and nucleon cross section via the massive spin-2 portal is
\be
\sigma_H=\frac{1}{32\pi}
\frac{\left(\kappa\kappa^\prime\right)^2}{9 M_{\mathrm{pl}}^4}\int_{-1}^{1} 
\frac{\d\cos{\theta}\, M^4 m_N^2}{[2m_N^2 v^2(1-\cos{\theta})+m_{\mathrm{eff}}^2]^2}\,,
\label{massive_amp}
\ee
and that of the massless spin-2 portal is
\be
\sigma_h=\frac{1}{32\pi}
\frac{\xi_r^4}{M_{\mathrm{pl}}^4}\int_{-1}^{1} 
\frac{\d\cos{\theta}\, M^4m_N^2}{\left[2m_N^2 v^2(1-\cos{\theta})\right]^2}\,,
\label{massless_amp}
\ee
We see that, due to the $M_{\mathrm{pl}}^4$ suppression (recall that we are considering $\xi_r \sim 1$), the massless portal cross section is too small to be observable in any DM direct detection experiment.
Therefore, we only need to consider the massive spin-2 portal, which can have a sizable cross section for a range of choices of $\beta$ and $\beta^\prime$.
Note that the massless cross section diverges at the forward limit $\cos{\theta}=1$. Since we are interested in DM direct detection, this can be regulated by the angle resolution of the experiment. The cut-off scattering angle $\Delta \theta$ is related to the recoil energy by $E_\text{recoil}=m_N v^2(1-\cos{\theta})=\frac{1}{2}m_N v^2\left(\Delta{\theta}\right)^2$. For instance, in the LZ experiment, the resolution of $E_\text{recoil}$ is $0.1$ keV for the Xenon target, which leads to $\left(\Delta{\theta}\right)^2$ about $10^{-3}$. However, for massive spin-2 portal that will be considered later, the difference between a resolution of $\left(\Delta{\theta}\right)^2=10^{-3}$ and a perfect resolution is negligible.

\section{The core-cusp problem} 

Now, we will see that, with suitable theory parameters, the model we propose can provide a solution to the core-cusp problem.
The core-cusp problem \cite{Moore:1994yx,Flores:1994gz,Navarro:1995iw,Navarro:1996gj} refers to the discrepancy between the observed DM density profile and that from simulations by collisionless cold dark matter: The simulations predict a steep increase of density at the center of a DM halo, while the observations of dwarf galaxies show a flat central density profile. It can be explained by the presence of DM self-interactions, as DM collisions can flatten the density profile at the center \cite{Spergel:1999mh,Tulin:2017ara,Kang:2020afi}. Thus, at work in short distances might be a light mediator, which in our case is filled by the massive spin-2 particle. 

In our bi-gravity model, if we take $\beta^\prime$ to be any $\mathcal{O}(1)$ value outside the range $[1-r,1+r]$, then we have $|\kappa^\prime|\sim 1/r$. Consequently, the effective fine structure constant for the non-relativistic DM self-scattering is given by $\alpha_{\rm{DM}}=\left({\kappa^{\prime }M}/{M_{\mathrm{pl}}}\right)^2/{4\pi}\sim \left({M}/{r M_{\mathrm{pl}}}\right)^2/{4\pi}$, which leads to a Yukawa potential $V(R)=-{\alpha_{\rm{DM}}}e^{-m_{\rm eff}R}/R$. On the other hand, since both dark matter and the massive spin-2 particles are associated with beyond the Standard Model new physics, we take the minimalist approach to assume that there is no sizable hierarchy between $M_f$ and the DM mass $M$. This yields $\alpha_{\rm{DM}}\sim 1/4\pi$. 

When the dark matter particle is much heavier than the massive spin-2 particle with $\alpha_{\rm{DM}}M/m_{\rm eff}\gg 1$, DM self-interactions are in the non-perturbative regime. With $Mv\gg m_{\rm eff}$, this is also a classical limit. Analytic results have been obtained for this non-perturbative regime at the classical limit by use of a screened Yukawa potential \cite{PhysRevLett.90.225002}. With these established, in order to solve the core-cusp problem for small scales (especially for the dwarf galaxy scales), 
it is found that when the effective coupling $\alpha_{\rm{DM}}\sim 0.01$ to 0.1, the mass of the mediator can have a range from $10^{-2}$\,GeV to $10^{-4}$\,GeV if the mass of the dark matter particle is from 100\,GeV to 10\,TeV \cite{Tulin:2013teo}. This allows us to pin down a parameter region for $m_{\rm eff}$ and $M$. While the massless graviton should couple to the SM sector at a strength of $M_{\mathrm{pl}}^{-1}$ to recover the Einstein gravity, the coupling for the massive spin-2 particle can in principle be different. However, a natural choice is that both of them are around the same order, considering the gravitational nature of the model. This can be effected by choosing $\beta=1-r/2$ for a given DM mass $M$ (thus a given $r=M/M_{\mathrm{pl}}$). This is also analogous to that of the electromagnetic and weak interactions where they originate from the same coupling but the masses of the mediators lead to dramatic differences in the interaction strength.

\begin{figure}[t]
        \centering
        \includegraphics[width=\linewidth]{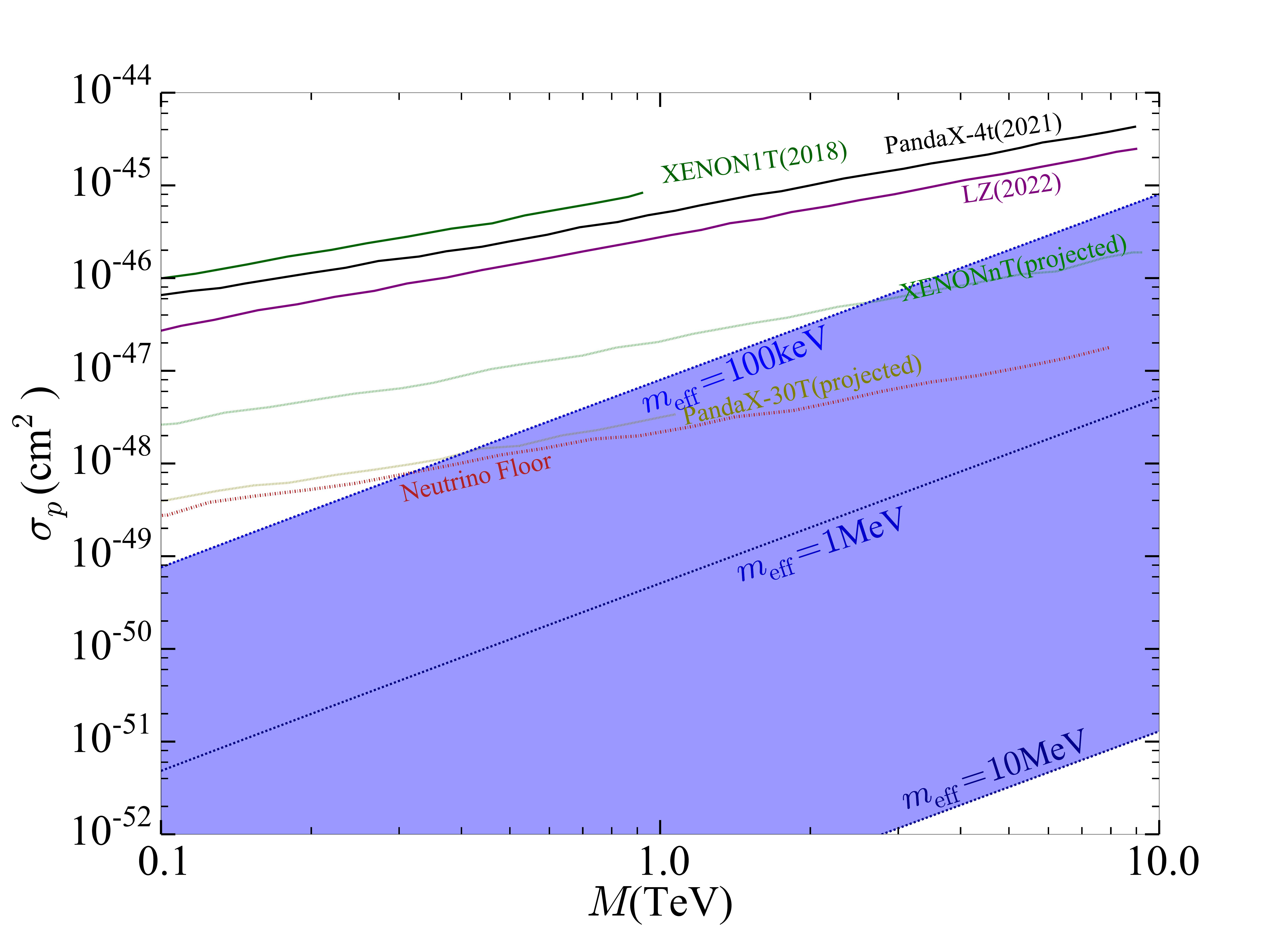}
\caption{\label{fig:cs_M}
Dark matter and proton elastic scattering cross section $\sigma_p$ versus mass of dark matter particle $M$ for various spin-2 mediator mass $m_{\mathrm{eff}}$. Also plotted are the sensitivities of the recent and projected dark matter direct detection experiments \cite{LZ:2022lsv,PandaX-4T:2021bab,XENON:2018voc,Aprile:2018dbl,Liu:2017drf,Aprile:2015uzo} as well as the neutrino background \cite{Billard:2013qya}.
}
\end{figure}

Concretely, for example, if we take $M=M_f=10^4$\,GeV, $\beta=1-r/2=1-10^{-14}/2$ and $\beta^\prime=0$ (thus we have $\kappa=1$, $\kappa^\prime=10^{14}$), for 10 \,TeV dark matter and $\alpha_{\rm{DM}}$ here, the mass of the spin-2 massive particle is about $10^{-4}$\,GeV (core-cusp problem solution), and then the cross section of dark matter and proton scattering is of order $10^{-47}\, \rm{cm}^2$. In Fig.\,\ref{fig:cs_M}, we plot how the cross section changes with the mass of the DM particle for a range of spin-2 mediator masses from $10^{-4}$\,GeV to $10^{-2}$\,GeV. We see that there is an observable window for this model in the next generation detectors, if the mass of the DM particle is from 1\,TeV to 10\,TeV and the mass of the massive spin-2 particle is from 100\,keV to 1\,MeV. 

\begin{figure}[t]
        \centering
        \includegraphics[width=\linewidth]{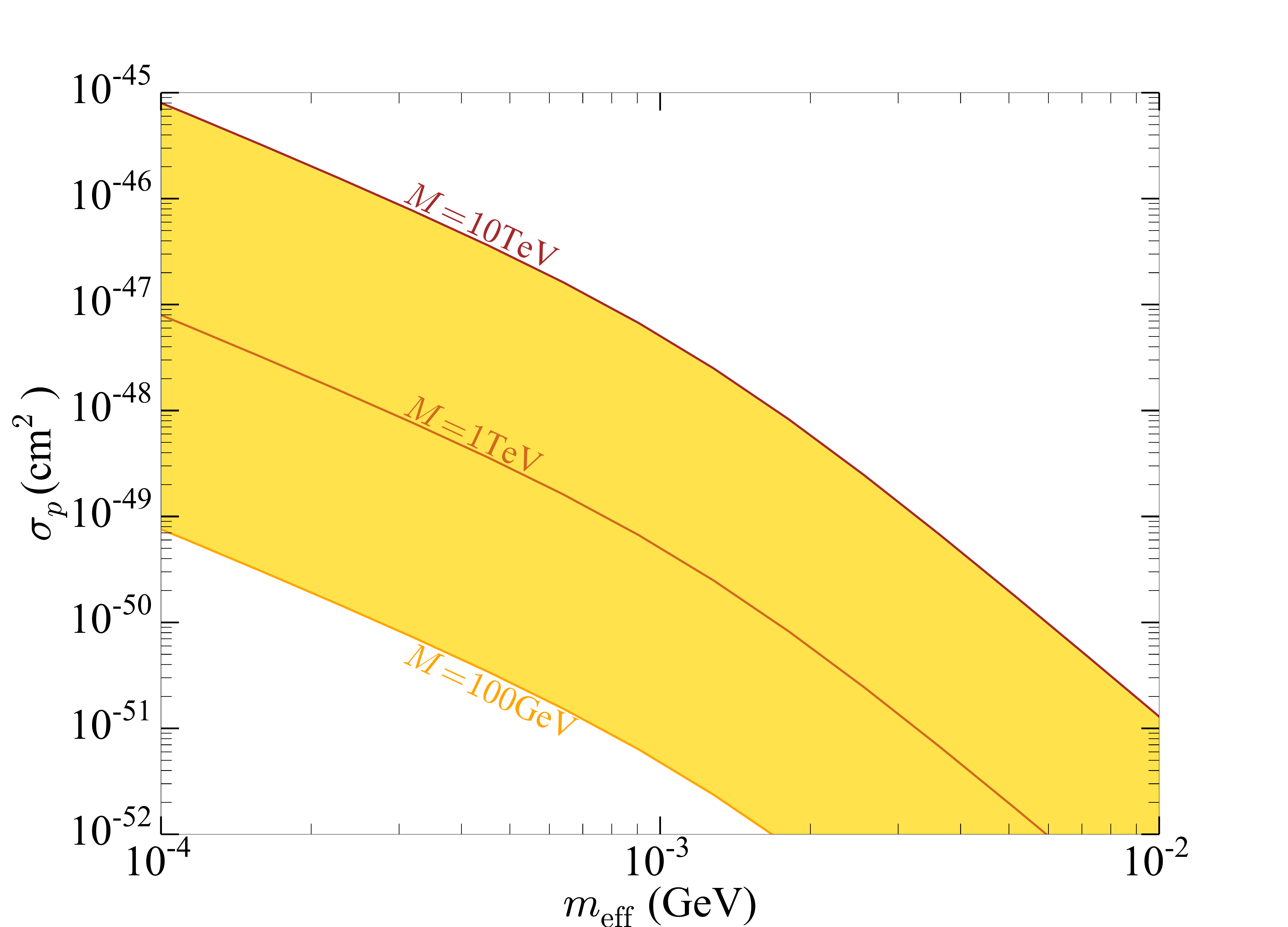}
\caption{\label{fig:cs_meff}
Dark matter and proton elastic scattering cross section $\sigma_p$ versus spin-2 mediator mass $m_{\mathrm{eff}}$ for various masses of the dark matter particle $M$.
}
\end{figure}

In the event of a direct detection of dark matter particles, we would be interested in probing the mass of the massive spin-2 particles. In Fig.\,\ref{fig:cs_meff}, we plot how the cross section changes with the spin-2 mediator mass when the mass of the DM particle varies from 100\,GeV to 10\,TeV. 

\section{Reheating temperature} 

An interesting prediction of this scenario is the reheating temperature. As the DM and SM interactions are too weak to reach a thermal equilibrium within a Hubble time, the dark matter relic abundance in this model is generated by a freeze-in mechanism \cite{Hall:2009bx,Bernal:2018qlk} at the reheating epoch, in which DM builds up its abundance slowly via interactions with the SM thermal bath. The production rate of dark matter in this model is approximately 
\ba
\dot{n}_{\rm{DM}}\simeq n^2_{\rm{SM}}\langle \sigma v \rangle \propto \frac{|\kappa\kappa^\prime|^2}{M_{\mathrm{pl}}^4}T^8\,,
\ea
where the SM number density $n_{\rm{SM}}\propto T^3$ and 
DM and SM conversion cross section $\sigma \propto  |\kappa\kappa^\prime|^2 T^2M_{\mathrm{pl}}^{-4}$. The dark matter abundance produced in a Hubble time $1/H$ is then
\be
Y_{\rm{DM}}= \frac{ n_{\rm{DM}}}{s} 
\propto \frac{n^2_{\rm{SM}}\langle \sigma v \rangle}{Hs}
\propto \frac{|\kappa\kappa^\prime|^2}{M_{\mathrm{pl}}^4} M_{\mathrm{pl}} T^3\,,
\ee
where $H\propto T^2/M_{\mathrm{pl}}$ and $s\propto T^3$ is the entropy density. The exact proportionality depends on the spin of dark matter \cite{Bernal:2018qlk}, with which we get
\be
\frac{\Omega_{\rm DM} h^2}{0.1}\approx 0.17\mu^{(s)}
\frac{|\kappa\kappa^\prime|^2}{10^{28}}
\frac{M}{10\, \text{TeV}}\left[\frac{T_{\mathrm{R}}}{10^{6}\, \rm{GeV}}\right]^3,
\ee
where the superscript in $\mu^{(s)}$ denotes the spin of the dark matter particle, with
$\mu^{(0)}=1$,  $\mu^{{(1/2)}}=6.1$,  $\mu^{(1)}=13.2$.
For our fiducial model, $M\sim 10$ TeV, $|\kappa\kappa^\prime|=2.4\times10^{14}$, 
we find that the reheating temperature $T_{\rm R}$ for a correct relic abundance observed is around $10^6\,{\rm GeV}$. Because $T_{\rm R}$ scales with $M^{1/3}$, this prediction remains highly robust across various DM masses.

\section{Gravity tests} 

With a nontrivial gravitational sector involved, one needs to check with the bounds from the general relativity tests. Our model easily passes all these tests. To see this, first note that since the massive spin-2 mode is very heavy, the linear approximation is valid in all the environments that are subject to the current gravity tests. So let us look at SM+SM $\to$ SM+SM scattering amplitudes. 
The massless portal scattering amplitude differs with the general relativity one by a factor of $\xi_r^2=(1+r^2)^{-1/2}$, where $r\leq 10^{-14}$. This is well below the current observational limits of the weak gravity tests\,\cite{Will:2014kxa}. The massive portal amplitude is exponentially suppressed by the Yukawa decay beyond the Compton wavelength of the massive spin-2 mode. For our model parameters considered in Fig.\,\ref{fig:cs_M}, the Compoton wavelength is shorter than $10^{-11}$\,m, which is again well within the observational limits of the graviton mass bounds or fifth force tests\,\cite{deRham:2016nuf, Will:2014kxa}.

\section{Summary}
We have investigated a bi-gravity model where the massless spin-2 particle gives rise to the ordinary Einstein gravity, while the massive spin-2 particle provides a sizable portal for the SM particles to interact with dark matter. (The massless spin-2 portal for the DM-SM coupling is of the usual gravitational strength and negligible in comparison at short range.) This is possible by using the unique, consistent composite metric couplings to matter, which utilizes the dRGT square-root construction. 
We have computed the cross sections of spin-independent elastic scattering between DM particles and protons, and confronted them with the experimental sensitivities of dark matter direct detection. Since the massive spin-2 portal also induces dark matter self-interactions, this model can account for the core-cusp problem. For it to be a solution to the problem, we find that there is an interesting region of the parameter space lying between the current experimental bounds and the sensitivity curves of the upcoming direct DM detection experiments. Of course, in the event of a successful DM detection, the argument can also be reversed to constrain the mass of the spin-2 particles. Specifically, in the viable parameter region, if the mass of DM particle is between 1\,TeV and 10\,TeV, the mass of the spin-2 portal particle needs to be about 100\,keV to 1\,MeV. Since a freeze-in mechanism is needed  to generate the DM relic abundance in this model, it also predicts a reheating temperature of $~10^6$ GeV for a large range of the DM mass.  

\section{Acknowledgments}
\begin{acknowledgments}
QC would like to thank Richard J.Hill for helpful discussions.
QC acknowledges postdoctorate fellowship supported by National Natural Science Foundation of China under grant No.\,12247103.
SYZ acknowledges support from the Fundamental Research Funds for the Central Universities under grant No.~WK2030000036, from the National Natural Science Foundation of China under grant No.~12075233 and 12247103, and from the National Key R\&D Program of China under grant No. 2022YFC220010.
\end{acknowledgments}

\bibliography{spin2portal}

\end{document}